\documentclass{aa}
\usepackage{graphicx}
\usepackage{txfonts}

\usepackage{natbib}
\bibpunct{(}{)}{;}{a}{}{,} 

\newcommand{\mabuls}{MaB$\mu$lS-2}
\renewcommand{\vec}{\boldsymbol}
\graphicspath{{./}{figures/}}

\usepackage{siunitx}
\usepackage{xcolor}
\usepackage{multirow}

\usepackage{mathtools}
\begin{document}
    \title{Euclid-Roman joint microlensing survey: early mass measurement, free floating planets and exomoons.}
    \subtitle{}

    \author{E. Bachelet \inst{1,7} 
    \and
    D. Specht \inst{2}
    \and
    M.Penny \inst{3}
    \and
    M. Hundertmark \inst{4}
    \and 
    S. Awiphan \inst{5}
    \and
    J.-P. Beaulieu \inst{6,7}
    \and
    M. Dominik \inst{8}
    \and
    E. Kerins \inst{2}
    \and
    D. Maoz \inst{9}
    \and
    E. Meade \inst{10}
    \and
    A. A. Nucita  \inst{11,12,13}
    \and
    R. Poleski \inst{14}
    \and
    C. Ranc \inst{4,7}
    \and
    J. Rhodes \inst{15}
    \and
    A. C. Robin \inst{16}
    }
    
    \institute{Las Cumbres Observatory, 6740 Cortona Drive, Suite 102,93117 Goleta, CA, USA\\
    \email{etibachelet@gmail.com}
    \and
    Jodrell Bank Centre for Astrophysics, The University of Manchester, Manchester M13 9PL, UK
    \and
    Department of Astronomy, The Ohio State University, 140 West 18th Avenue, Columbus, OH 43210, USA
    \and
     Astronomisches Rechen-Institut, Zentrum f{\"u}r Astronomie der Universit{\"a}t Heidelberg (ZAH), 69120 Heidelberg, Germany
    \and
    National Astronomical Research Institute of Thailand, 260 Moo 4, T. Donkaew, A. Maerim, Chiangmai 50180, Thailand
    \and
    School of Physical Sciences, University of Tasmania, Private Bag 37 Hobart, Tasmania 7001 Australia
    \and
    Sorbonne Universit\'es, CNRS, UMR 7095, Institut d'Astrophysique de Paris, 98 bis bd Arago, 75014 Paris, France
    \and
    University of St Andrews, Centre for Exoplanet Science, SUPA School of Physics \& Astronomy, North Haugh, St Andrews, KY16 9SS, United Kingdom
    \and
     School of Physics and Astronomy, Tel Aviv University, Tel Aviv 69978, Israel
    \and
    Department of Physics, The University of Texas at Dallas, 800 W Campbell Road, Richardson, TX 75080, USA
    \and
    Department of Mathematics and Physics, University of Salento, via per Arnesano, CP-73100, Lecce, Italy
    \and
    INFN, Sezione di Lecce, Via per Arnesano, CP-193, I-73100 Lecce, Italy
    \and
    INAF, Sezione di Lecce, Via per Arnesano, CP-193, I-73100 Lecce, Italy
    \and
    Astronomical Observatory, University of Warsaw, Al. Ujazdowskie 4, 00-478 Warsaw, Poland
    \and
    Jet Propulsion Laboratory, California Institute of Technology, 4800 Oak Grove Drive, Pasadena, CA, 91109, USA
    \and
    Institut UTINAM CNRS UMR6213, Universit\'{e} Bourgogne Franche-Comt\'{e}, OSU THETA Franche-Comt\'{e} Bourgogne, Observatoire de Besan\c{c}on, BP1615, 25010 Besan\c{c}on Cedex, France
}
   \date{Received ??; accepted ??}

\abstract
{As the {\sc Kepler} mission has done for hot exoplanets, the ESA {\sc Euclid} and NASA {\sc Roman} missions have the potential to create a breakthrough in our understanding of the demographics of cool exoplanets, including unbound, or "free-floating", planets (FFPs). {\sc Roman} will dedicate part of its core survey program to the detection of cool exoplanets via microlensing, while {\sc Euclid} may undertake a microlensing program as an ancillary science goal. In this study, we demonstrate the complementarity of the two missions and propose two joint-surveys to better constrain the mass and distance of microlensing events. We first demonstrate that an early brief {\sc Euclid} survey ($\sim 7$ h) of the {\sc Roman} microlensing fields will allow the measurement of at least 30\% of the events' relative proper motions $\mu_{rel}$ and 42\% of the lens magnitudes. This survey would place strong constraints on the mass and distance on thousands of microlensing events observed by {\sc Roman} just after the first year of observation. Then, we study the potential that simultaneous observations by {\sc Roman} and {\sc Euclid} to enable the measurement of the microlensing parallax for the shortest microlensing events and, ultimately, obtain a direct measurement of the masses, distances and transverse motions of FFPs. Using detailed simulations of the joint detection yield we show that within one year {\sc Roman}-{\sc Euclid} observations will be at least an order of magnitude more sensitive than current ground-based measurements. The recent tentative detection of an excess of short-duration events by the {\sc OGLE} survey is consistent with a scenario of up to 10 Earth-mass FFPs per Galactic star. For such a scenario a joint {\sc Roman}-{\sc Euclid} campaign should detect around 130 FFP events within a year, including 110 with measured parallax that strongly constrain the FFP mass, and around 30 FFP events with direct mass and distance measurements. The ability of the joint survey to completely break the microlens mass-distance-velocity degeneracy for a significant subset of events provides a unique opportunity to verify unambiguously the FFP hypothesis or else place abundance limits for FFPs between Earth and Jupiter masses that are up to two orders of magnitude stronger than provided by ground-based surveys. Finally, we study the capabilities of the joint survey to enhance the detection and charcterization of exomoons, and found that it could lead to the detection of the first exomoon.}



\titlerunning{Euclid-Roman joint microlensing survey}
\authorrunning{Bachelet et al.}
\maketitle

\section{Introduction} \label{sec:intro}

Our understanding of exoplanet demographics has progressed enormously thanks to advances in exoplanet detection techniques that have led to the rapid growth of confirmed candidates\footnote{\label{exo-eu} According to The Exoplanet Encyclopedia (\label{exo-eu}\protect{\url{http://www.exoplanet.eu}}), accessed 3rd January 2022.}. Some 3,484 of the 4,909 confirmed exoplanets discovered to date have come from primary transit observations, with around 2,300 of these coming from the {\sc Kepler} space mission alone\textsuperscript{\ref{exo-eu}}. The ability to observe from space has enabled robust, large-scale, regular and homogeneous photometric observations that are crucial for extending exoplanet sensitivity down to the Earth-size, or even sub-Earth-size, regime. While the list of confirmed exoplanets now extends into the thousands, our knowledge of their demographics still suffers from severe incompleteness. Many of the current candidates are relatively massive and on short-period orbits compared to planets in our own solar system. We have incomplete knowledge about the full extent of multi-planet systems, particularly for those systems that, like our own solar system, host planets beyond the ice line. 
Understanding the planet population beyond the ice line is critical for understanding planet formation. Population synthesis simulations show that, unlike massive planets, low-mass planets that form beyond the ice line are not strongly affected by migration as their accretion timescales are short compared to the migration timescale \citep{Mordasini2018}. This implies that cool, low-mass planets provide key tracers of \emph{in-situ} planet formation. Studying the present-day demographics of cool, low-mass planets can, uniquely, provide a direct test of planet formation models that is largely unaffected by complex planet migration histories.

Gravitational microlensing remains the only available method able to access the cool, low-mass exoplanet regime. Ground-based microlensing surveys have so far detected over 100 planets\textsuperscript{\ref{exo-eu}}.
Microlensing data have been used to show that the distribution of planet-to-host mass ratios exhibits a universal profile and therefore may be a more primary diagnostic of planet formation than the planet mass function \citep{Suzuki2016,2018ApJ...856L..28P}. Ground-based microlensing data have also tentatively indicated the existence of a significant population of planets that are isolated or very distant (i.e., $\ge15$ AU) from any host star \citep{Sumi2011,Mroz2017,2019A&A...622A.201O}, often referred to as free-floating planets (FFPs). The first FFPs candidates have been discovered two decades ago in star forming regions \citep{Oasa1999,Zapatero2000,Luhman2005,Burgess2009,Marsh2010} and are predicted by planet formation models \citep{Veras2012,Ma2016}. The existence of FFPs is also implicated by infrared surveys of nearby star-forming regions \citep{2017ApJ...842...65Z}. In particular, \citet{MiretRoig2021} recently found a $\sim 5$ \% relative abundance of FFPs (with a mass range of $4-13 M_{\rm Jup}$) in their survey of the Upper Scorpius and Ophiuchus regions. However, their Galactic abundance and typical mass scale remains uncertain. Because it does not rely on the measurement of the light emitted by the lens, microlensing is a powerful technique to discover these objects \citep{Sumi2011,Mroz2017,Mroz2019}. Whether FFPs form in isolation or exist as a consequence of being ejected from their hosts through dynamical exchanges with other planets remains unknown. Clearly, it is important to pin down the nature and abundance of FFPs in order to answer this question and to inform planet formation models. Microlensing is the only method available to study FFPs over Galactic distance scales.

The primary observable, the Einstein radius crossing time $t_E$ not only depends on the lens mass and distance from the observer, but also on the relative proper motion and relative parallax between lens and source \citep{Gould2000}. The characterization of the lens mass and distance is therefore challenging and requires the measurements of some extra parameters, such as the microlensing parallax \citep{Gould1994}, finite-source effects in the lightcurve \citep{Witt1994}, the source and lens relative proper motion and/or the lens flux, see for example \citet{Vandourou2020}. So far, it has be done only for about half of the bound microlensing planets. One example is OGLE-2015-BLG-0966Lb \citep{Street2016}, a lens system consisting of a 0.4~M$_{\odot}$ M~dwarf, orbited by a cold Neptune. It was simultaneously observed from the ground and with the {\sc Spitzer} Space Telescope, to obtain a direct planet mass measurement with a $10\%$ precision. Ground and space-based observations have also been used to directly image the lensing host star, several years after an event has occurred. The host lens flux can be then used to convert the planet-to-host mass ratio, obtained from the microlensing light curve, to a planet mass \citep{Yee2015,Beaulieu2018}.

Due to the intrinsic rarity of microlensing, surveys must target highly crowded stellar fields toward the Galactic bulge. As a result, their sensitivity is severely affected by the combination of stellar crowding effects and seeing limitations caused by Earth's atmosphere. As with the transit method, substantial sensitivity improvements can therefore be achieved by observing from space \citep{Bennett1996,Penny2013,Penny2019}. 
For microlensing, the primary signal is a time-dependent magnification of a background source whose light is deflected by the gravitational field of a foreground object. Because the gravitational lensing cross section (the Einstein ring radius) of the lens scales with the square root of its mass, the event duration is a function of the lens mass, up to several years for a $\sim 10 M_\odot$ isolated mass black hole \citep{Wyrzykowski2016} and down to few hours for a FFP \citep{Mroz2020}. If the angular Einstein ring radius $\theta_E$ is small compared to the angular size of the background source star $\theta_*$, the magnification averaged over the face of the source may be too small to be detected. A major benefit of observing from space is the ability to access large numbers of resolved (i.e., not blended with other stars) main sequence stars, as from the ground most of the non-blended stars are the larger bulge giants. The more-compact main-sequence stars extend microlensing sensitivity down to around the mass of the Moon \citep{Dominik2007,Penny2019}.

In the next decade, both NASA and ESA will be launching facilities that could undertake high precision exoplanet microlensing surveys. The NASA {\sc Nancy Grace Roman Space Telescope} (formerly {\sc WFIRST} and hereafter referred to as {\sc Roman} -- \citealt{Spergel2015}) core missions include a dark energy survey and a substantial microlensing program. The ESA {\sc Euclid} mission's \citep{Laureijs2011} primary objective is to constrain the nature of dark energy, although the original mission definition included also the possibility of undertaking a microlensing survey as an additional goal \citep{Penny2013}. Both telescopes are scheduled to be located in halo orbits at the Earth-Sun L2 point, with a potential separation between them of up to 600,000~km. Microlensing simulations have shown that both telescopes are ideal for microlensing, and have the potential to detect thousands of planets across a broad range of masses and semi-major axes \citep{Penny2013,Penny2019}. \citet{Johnson2020} predicts that {\sc Roman} alone will detect $\sim$ 250 FFPs, including $\sim 60$ planets less massive than the Earth, in good agreement with the estimation of \citet{Ban2016}. \citet{Johnson2020} also forecasts that a significant fraction of the FFP events detected by {\sc Roman} will display finite-source effects. This implies that $\theta_E$ will be measured for most of these events, because $\theta_*$ will be almost systematically known for events detected by {\sc Roman}. On the other side, \citet{Bachelet2019} shows that the microlensing parallax will not be detected by {\sc Roman} alone for events with $t_E \le 3$ days, i.e. outside of the FFP regime. However, the authors show that \textit{simultaneous} observation from the {\sc Euclid} mission would constrain the microlensing parallax down to the FFP regime, if the separation between the two mission is sufficiently large. This was confirmed by the independent analysis of \citet{Ban2020}, that demonstrates that the parallax will be measurable for a large fraction of events due to FFP.  

In this paper, we argue that the {\sc Roman} and {\sc Euclid} missions should coordinate their observations to undertake a combined survey to enhance the characterization of the lensing systems. The paper is layed out as follows. In Section~\ref{sec:timelines}, we overview the {\sc Roman} and {\sc Euclid} missions and their respective survey timelines, pointing out the options for a simultaneous observing programme. In Section~\ref{sec:pmflux}, we visit the theory behind precision mass measurements of microlensing events. In Section~\ref{sec:early}, we consider the potential of an early {\sc Euclid} survey of the {\sc Roman} microlensing fields for the measurements of the lens fluxes as well as the relative proper motions. In Section~\ref{simul}, we present  a statistical study that details the unique constraints that offer a {\sc Euclid} and {\sc Roman} joint-survey on the detection and characterization of microlensing events due to FFP. Section~\ref{sec:exomoons} considers the ability of a joint survey to detect exomoons. We conclude with a summary in Section~\ref{sec:sum}.

\section{{\sc Roman} and {\sc Euclid} missions} \label{sec:timelines}
\subsection{Timelines}

{\sc Roman} is a NASA flagship mission scheduled for launch in late 2025.  {\sc Roman} has three primary science goals: dark energy investigations using galaxy clustering, weak gravitational lensing, and Type Ia supernovae; exoplanet demographics via microlensing; and infrared surveys for a wide range of astrophysics applications, executed via a competitive General Observer (GO) program \citep{Spergel2015}. Additionally, {\sc Roman} will carry a Coronagraph Instrument (CGI) technology demonstration that will premiere, in space, a number of new techniques that will advance the state-of-the-art in coronagraphy by three orders of magnitude in star/planet contrast ratio. Additional surveys are possible within the primary mission, and via the GO program and a possible 5-year extension.

{\sc Euclid} is an ESA-led medium-class mission with significant technical and scientific contributions from NASA.  {\sc Euclid} is designed to map the universe with two distinct probes, galaxy clustering and weak gravitational lensing, in order to study the nature of dark matter and dark energy \citep{Laureijs2011}.  {\sc Euclid} will perform these surveys with two instruments: VIS will do visible photometry in a single, very-wide band, and the NISP instrument will perform near-infrared photometry and grism spectroscopy between $1-2\mu$m. {\sc Euclid} is scheduled to launch in late 2022 for a 6-year primary mission. As the Euclid mission progresses, there will be increasing  gaps in the cosmology survey that could be filled with microlensing observations. The  {\sc Euclid} NISP instrument is expected to be fully functional after the 6-year primary survey and therefore microlensing observations using NISP could be viable. The VIS CCDs will likely suffer from increasing charge-transfer-efficiency degradation throughout the mission, with a commensurate degradation of the VIS capabilities. While this may impact their suitability for weak lensing measurements, it's unlikely that increased CTE would dramatically degrade VIS's ability to perform microlensing photometry (a full analysis of the effects of detector degradation on VIS microlensing measurements after the Euclid prime mission is beyond the scope of this paper).

\subsection{Relevant Instruments}

The {\sc Roman} Wide Field Instrument (WFI) consists of 18 4k$\times$4k Hawaii H4RG-10 HgCdTe detectors with a pixel scale of $0.11''$, giving a 0.28 deg$^2$ field of view~\citep{Spergel2015}.  In addition to a grism and prism, the instrument will have six broad band filters with central wavelengths spanning from $0.62$~$\mu$m to $2.13$~$\mu$m ($R062$, $Z087$, $Y106$, $J129$, $H158$, $F184$, $K213$) as well as an extremely wide $0.92$--$2.00$~$\mu$m filter, $W146$, that will be used by the Galactic exoplanet survey.  WFI delivers diffraction limited imaging for the 2.4-m {\rm Roman} telescope in all but the shortest wavelength filter.  {\sc Roman} is able to slew to and settle on an adjacent field in about 60 seconds using reaction wheels for maneuvers. 

{\sc Euclid} uses a $1.2$-m telescope that delivers light to two instruments that simultaneously observe the same field through a dichroic element~\citep{Laureijs2011}.  The VIS instrument has 36 4k$\times$4k e2v CCD273-84 CCD detectors with $0.10''$ pixels, giving a 0.47 deg$^2$ field of view.  VIS provides diffraction limited imaging in an extremely wide optical light bandpass of $0.55$--$0.90$~$\mu$m~\citep{Cropper2018}.  The NISP instrument images the same field as VIS with 16 H2RG-18 HgCdTe detectors with a pixel scale of $0.30''$~\citep{Maciaszek2016}.  NISP can select from four grisms or three broadband filters with central wavelengths of $1.05$, $1.37$, and $1.77$~$\mu$m. {\sc Euclid}'s primary observing modes will allow VIS imaging to be captured simultaneously with NISP grism observations, but in principle it should be possible for VIS and NISP to both perform imaging observations simultaneously.  Similarly, the number of VIS and NISP exposure time modes that are commissioned may be limited.  {\sc Euclid} uses cold gas thrusters to maneuver, which results in relatively long ${\sim}350$ second slew and settle times to move to an adjacent field~\citep{GomezAlvarez2018}.

\subsection{Planned Bulge Surveys}

 {\sc Roman} will conduct the {\sc Roman} Galactic Exoplanet Survey, which will nominally consist of 6 ``seasons'' of continuous observations lasting 72 days each~\citep{Penny2019}.  The seasons will be spaced six months apart, surrounding the vernal and autumnal equinoxes when {\sc Roman} can point toward the bulge.  During this time, {\sc Roman} will observe ${\sim} 2$~deg$^2$ of the Galactic bulge every 15 minutes in its wide $W146$ filter, and at least once every twelve hours in at least one of the broad-band filters, currently planned to be $Z087$.
 
 {\sc Euclid} will not conduct bulge observations as part of its primary mission, but we consider here two scenarios where bulge observations would be minimally disruptive to the {\sc Euclid} primary mission: a short observing campaign of pre-imaging of the {\sc Roman} bulge fields shortly after launch, and more extensive observations simultaneously with {\sc Roman} late in the {\sc Euclid} mission. {\sc Euclid} can point toward the bulge in a narrower ${\sim}30$ day window while maintaining thermal stability that is fully enclosed within the 72~day {\sc Roman} seasons.

\subsection{Orbital elements}

Both telescopes will orbit around the L2 Sun-Earth Lagrangian point. While the exact orbital elements are not known yet, it is safe to assume that the orbits will be comparable to the current orbit of the {\sc Gaia} telescope, with an orbital radius $R\sim300,000$ km and a period of $\sim$ 180 days. Using these parameters, \citet{Bachelet2019} have shown that simultaneous observations from the two telescopes unlock the parallax measurement down to the FFP regime, if the telescope's separation is at least $\ge 100,000$ km. We use these parameters for the rest of this work.

\section{Measurement of the masses and distances of microlenses} \label{sec:pmflux}
A microlensing event occurs when a lens object at a distance $D_L$ crosses the line of sight between the observer and a source at a further distance $D_S$. When the angular distance between the source and the lens is sufficiently small (a few $\theta_E$, see below), the gravity field of the lens modifies the pathway of the photons and creates several images of the source. The images are distributed around the lens and separated by few Einstein ring radius $\theta_E$ \citep{Gould2000},
 \begin{equation}
     \theta_{\rm E} = \sqrt{\frac{4GM_L\pi_{rel}}{c^2}}\approx88 ~\mu\mbox{as} ~ 
     \left( \frac{M_{\rm L}}{M_{\rm Jup}} \frac{\pi_{rel}}{\rm mas}\right)^{1/2}
\end{equation}
where
\begin{equation}
    \pi_{\rm rel} = \biggl(\frac{1}{D_L}-\frac{1}{D_S}\biggr)~ {\rm au~ kpc^{-1}}
\end{equation}
and $M_L$ is the lens mass, G is the gravitational constant and c the speed of light in vacuum. As long as no finite source effects are measured, there is only a 
 single measurable parameter that depends on the lens mass, namely the Einstein radius crossing time,
 \begin{equation}
     t_{\rm E} = \frac{\theta_{\rm E}}{\mu_{\rm rel}} = 1~\mbox{day}~\left( \frac{M_{\rm L}}{M_{\rm Jup}} \right)^{1/2} \left( \frac{\pi_{rel}\mbox{10 kpc}}{au} \right)^{-1/2}\left( \frac{\mu_{\rm rel}}{\mbox{10 mas yr}^{-1}} \right)^{-1}.
     \label{eq:te}
 \end{equation}
Worse still, $t_{\rm E}$ also depends on the relative lens--source proper motion, $\mu_{\rm rel}$, giving rise to a three-parameter degeneracy between $M_{\rm L}$, $\mu_{\rm rel}$ and $\pi_{\rm rel}$.
The physical mass and distance to the lens system can be estimated via a Bayesian analysis using a Galactic model to define the priors on $\pi_{rel}$ and $\mu_{\rm rel}$ \citep{Han1995,Han2003,Bennett2014}. Since these priors are rather broad towards the Galactic bulge, whilst the planet-to-host mass ratio is often well determined, absolute values of physical parameters may not be well constrained.
Fortunately, there are a number of ways to break the microlensing parameter degeneracy through combinations of high-precision photometry, high resolution imaging follow-up, and simultaneous observations from well-separated observatories. We briefly review such methods that could be employed by {\sc Roman}.

\subsection{Mass-distance relation from finite source effects}

The Einstein radius, $\theta_{\rm E}$, can be measured if a light curve shows evidence of finite source size effects, e.g., when the source approaches caustics of the lensing system. For single lens, this occurs when the angular separation separation $\beta$ between the source and the lens becomes comparable to the angular size of the source star $\theta_*$, i.e. $\beta \lesssim \theta_*$. This is also a common situation for microlensing events that involve a planetary lens, and it results in a differential magnification across the source face that is observable as a deviation of the microlensing light curve from the point-source model. The effect essentially allows the source angular size to be used as a ``standard ruler'' to measure $\theta_{\rm E}$ via
\begin{equation}
    \rho_* = \theta_* / \theta_{\rm E},
  \label{star-size}
\end{equation}
where $\rho_*$ is the angular size of the background source star in units of $\theta_{\rm E}$. $\rho_*$ is a parameter that can be determined from a fit of a light curve exhibiting finite source effects. The source angular size, $\theta_*$, can be determined reliably via the infrared surface brightness relation \citep{Yooo2004,2013ApJ...771...40B}, determining $\theta_{\rm E}$ through   Eq~(\ref{star-size}). This in turn yields a measurement of $\mu_{\rm rel}$ through Eq~(\ref{eq:te}). Once $\theta_{\rm E}$ has been measured, the microlens degeneracy reduces to a mass-distance relation
\begin{equation} 
 M_{\rm L} =  \frac{ \theta_{\rm E}^2 }{ \kappa~\pi_{ \rm rel}},
  \label{mass-pie}
\end{equation}
with $\kappa \equiv 8.144~\mbox{mas}~M_\odot^{-1}$.

{\sc Roman} will provide dense, high-precision, photometry on planetary microlensing events. Since a large fraction of these systems are expected to exhibit finite-source effects (either from low-mass FFP light curves, or from bound planets detected via caustic crossings by the source), this type of mass-distance relation will be obtained routinely.

\subsection{Mass-distance relation from lens flux}
With high-enough angular resolution observations, it is possible to disentangle the aligned source and lens stars from field stars at the subarcsec scale. Modeling of the photometric light curve  allows the flux of the source star to be estimated accurately. In principle, one can then measure the excess flux aligned with the source star, and try to determine if some or all of this excess flux can be attributed the planetary host star. A complication, at this point, is that there might be contaminants---chance-aligned field stars, a companion to the source star, or another stellar companion to the host. These possibilities must be evaluated via a Bayesian analysis \citep{2019arXiv191011448K}. The resulting lens magnitude $m_{\rm L}(\lambda)$ can be combined with an empirical mass-luminosity  relation \citep{2000A&A...364..217D} or with stellar isochrones \citep{2008A&A...484..815B} 
to get a mass-distance constraint of the form 
\begin{equation}\label{eq:HR}
 m_{\rm L}(\lambda) = 10 + 5\log(D_{\rm L}/\mbox{kpc}) + A_{\rm L}(\lambda) + M_{ \rm isochrone}(\lambda, M_{\rm L}, \mbox{age}, [\mbox{Fe/H}]), 
 \end{equation}
where $M_{ \rm isochrone}$ is the absolute magnitude of the star at wavelength $\lambda$. If several bands can be used in concert, the constraints on the mass and distance of the lens, as well as the extinction along the line of sight, become stronger \citep{Batista2015}. And as soon as the lens and the source are sufficiently separated (i.e. when the separation is $\gtrsim 0.5$ FWHM), generally several years after the event peak, the exact nature of the lens is known with very high precision \citep{Vandourou2020,Bhattacharya2021}. This is also the case for stellar remnants lenses, because the high resolution images can rule-out the main-sequence lens scenario \citep{Blackman2021}. Of course, this measurement is extremely challenging, if not impossible, for the faintest components, such as FFPs or stellar remnants lenses.

\subsection{Mass-distance relations from lens-source relative proper motion}

The lens-source relative proper motion introduced in Eq~(\ref{eq:te}) is typically of the order of $\sim 5~{\rm mas~yr^{-1}}$ and generally in the range $\sim1-10~{\rm mas~yr^{-1}}$.
With the very stable point spread function (PSF) of {\sc Roman}, it is possible to measure the centroid variations even if the source and the lens are not fully resolved from each other, to constrain their flux ratio (as described in the previous sub-section) and relative proper motion. Observing the microlensing event in several bands significantly increases the precision of the method, often referred as "color-dependent centroid shift". This has already been achieved, for example using the \textsc{Hubble Space Telescope} \citep{Bhattacharya2018}. The measurement of the proper motion, especially coupled with the measurement of the lens flux, provides strong constraints on microlensing models. This is particularly useful for breaking fundamental degeneracies that can arise in the light curve modeling, especially the ``ecliptic degeneracy'' \citep{Skowron2011}. Again, this method is almost impossible to use for the faintest lenses of the Milky Way.

\subsection{Mass-distance relations from microlensing parallax}\label{sec:para}

While measurements of $t_{\rm E}$ and $\mu_{\rm rel}$ (or, equivalently, $\theta_{\rm E}$) break some of the microlensing parameter degeneracy, we require an additional mass-sensitive measurement to fully resolve the degeneracy. The reward is not just the lens mass but also its distance and transverse velocity, both of which are of physical interest. The relative lens-source parallax $\pi_{\rm rel}$ introduced in Eq~(\ref{mass-pie}) can be measured if  $t_{\rm E}$ is long enough to observe a subtle shift in the photometric light curve due to the Earth's orbital motion (the annual parallax -- \citealt{Gould2004}), or if two observatories observe the event simultaneously (the satellite parallax --  \citealt{Refsdal1966}).  In this case, microlensing parallax vector is approximately given by \citep{CalchiNovati2015}: 

\begin{equation}
\boldsymbol{ \pi_{\rm E}} = {{\rm au}\over{ D_\perp}}  \left({{\Delta t_0}\over{t_{\rm E}}},\Delta u_0\right),
\label{eq:separation}
\end{equation}
where ${ D_\perp}$ is the projected separation between the two observatories along the direction of the event, $\Delta u_0$ and $\Delta t_0$ are the differences in impact parameter and epoch of maximum magnification recorded by the two observatories, respectively.  This leads again to the mass-distance relation of Eq~(\ref{mass-pie}), but recast in the form
\begin{equation}
M_{\rm L} = {{\theta_{\rm E}}\over{\kappa\pi_{\rm E}}}, ~~\mbox{where}~\pi_{\rm E}  = \frac{\pi_{\rm rel}}{\theta_{\rm E}}.
\label{piE}
\end{equation}
The switch from $\pi_{\rm rel}$ to $\pi_{\rm E}$ highlights that the key to detecting microlensing parallax is to obtain measurements of the light curve from positions that are separated over distances that, when projected onto the lens plane, span a sufficient fraction of the lens Einstein radius. 

Recently, it has been demonstrated that the {\sc Roman} observations from the Earth-Sun L2 point will be sufficient to detect parallax for microlensing events with $t_{\rm E}\gtrsim5$ days, due to the orbit of the L2 point around the Sun \citep{Bachelet2019}. This will ensure strong constraints on the mass and distance of the $\sim 1,500$ bound planets expected to be detected by {\sc Roman} \citep{Penny2019}.
FFP microlensing events, however, are generally expected to have timescales of less than a few days. Measurements of parallax for such short $t_{\rm E}$ events from surveys such as {\sc Roman} or {\sc Euclid} operating alone will be extremely challenging. \citet{Hamolli2016} find a relatively high efficiency for parallax measurement when events can be detected down to peak magnifications of 1.001, irrespective of signal to noise. \citet{Hamolli2016} also determine parallax detection based on a comparison of the model residuals for the same event, with and without parallax, rather than between the best-fit parallax model and a best-fitting model without parallax.  Without very high signal-to-noise observations, we consider such a low detection threshold to be unreliable for secure false-positive rejection. For more robust selection criteria based on an event's signal-to-noise ratio, and a comparison of best-fit models with and without parallax, we would not expect either survey to be highly efficient to FFP parallax detection when observing alone.

However, {\em simultaneous}\/ observations from {\sc Euclid} and {\sc Roman} are ideal to measure the parallax for such events \citep{Bachelet2019}. This is of particular interest because, as noted, FFPs are expected to produce short events with time scales of hours to days (Eq~\ref{eq:te}). For a typical FFP lens of 1 $M_{Jup}$, located at $D_L=4$ kpc and a source at $D_S=8$ kpc, the projected Einstein ring radius $\tilde{r_E}=\theta_E / \pi_{rel}$  \citep{Gould2000} is about 0.25 au. For an orbital radius around L2 of $\sim$ 300,000 km, the maximum separation between {\sc Euclid} and {\sc Roman} can be of the order of $\sim0.005$ au, a range that allow the measurement of the parallax of FFP events. Rewriting Equation~\ref{eq:separation} gives:
\begin{equation}
\Delta t_0 = {{D_\perp\pi_{rel}}\over{au ~ \mu_{rel}}}
\end{equation}
and
\begin{equation}
\Delta u_0 = 0.35~\frac{D_\perp}{\rm au}\sqrt{{{\pi_{rel}}\over{\rm mas}} {{M_\odot}\over{M_L}}}
\end{equation}
Assuming $\mu_{rel}=4$ mas/yr, the previous configuration leads to $\Delta t_0\sim1$ hour and $\Delta u_0\sim0.02$, offsets that should be measurable given the cadence and photometric precision of the surveys. Similarly, \citet{Yee2013} and \citet{Bennett2018} described how simultaneous observation from {\sc Roman} and ground-based telescopes would constrain the mass of FFP lenses. However, it is inevitable that the lower cadence and photometric precision of observation collected from the ground make these measurements more challenging.

\section{Direct Planet Host Mass Measurements with Early {\sc Euclid} observations}
\label{sec:early}
In the case of the lens host a secondary object, the mass ratio $q$ and the projected separation $s$ has to be accounted in the modeling of the microlensing lightcurve. These parameters will be precisely measured for planets down to the mass of Mars by the {\sc Roman} microlensing survey \citep{Penny2019}. However, as previously presented, additional measurements are required to accurately derive the host and planet masses. In this section, we will demonstrate that an early epoch of imaging of the {\sc Roman} microlensing fields can potentially improve lens mass measurements, including those for planet hosts, by increasing the time baseline over which the lens and source separation can be observed \citep{Yee2014}. Presently, launch dates of 2022 and 2026 are expected for {\sc Euclid} and {\sc Roman}, respectively, meaning that an epoch of imaging early in {\sc Euclid}'s mission could extend the baseline by up to four years relative to the {\sc Roman} microlensing survey's expected 3.5-4.5 year baseline. \citet{Bennett2007} estimate that the precision of both lens flux and source-lens relative proper motion measurements will scale inversely with the square root of the number of photons ($N^{-1/2}$) but as the inverse cube of the lens-source separation, or proper motion baseline ($\Delta t^{-3}$). Therefore, a relatively short program to observe the {\sc Roman} microlensing fields using a few hours of exposure time with {\sc Euclid} could provide competitive, if not superior lens mass measurements for at least a subset of {\sc Roman}'s expected ${\sim}1,500$ exoplanet discoveries, despite the almost ${\sim}100$ hours of exposure time {\sc Roman} will collect of each field each season. In this section we use a simulation of Euclid images to demonstrate this possibility.

\subsection{Description of simulations}

\begin{figure*}[h]
    \centering
    \includegraphics[width=0.45\textwidth]{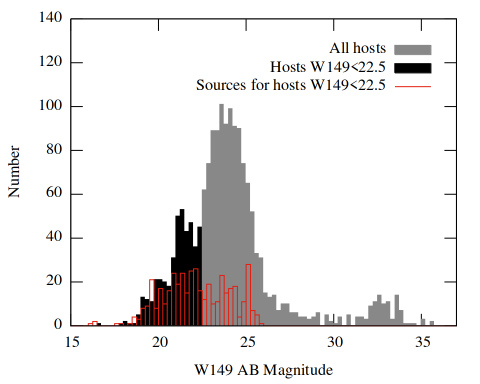}
    \includegraphics[width=0.45\textwidth]{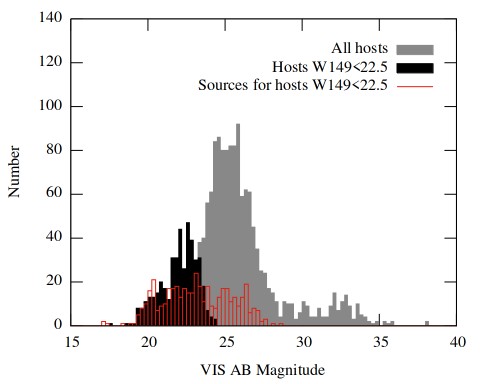}
    \caption{Distribution of host star (lens) magnitudes from a sample of simulated planetary microlensing events injected in the {\sc Euclid} image. The grey histogram shows the $W149$ (left) and $VIS$ (right) magnitude distribution of all hosts in a sample of $1691$ planetary microlensing events drawn from simulations presented in \citet{Penny2019}, and the black histogram shows the distribution of $432$ events brighter than $W149<22.5$, which were injected into our simulated image. The red histogram shows the distribution of  magnitudes for the sources of the $432$ events with bright lenses.}
    \label{figure:ImgSimInput}
\end{figure*}

We simulated an early {\sc Euclid} image of a star field containing microlensing events that Roman will observe. We used the image simulation component of the {\sc gulls} microlensing simulator~\citep{Penny2013,Penny2019}. A realization of a $1.83$'$\times 1.83$' starfield was produced by drawing stars from the {\bf BGM1307} version of the Besan{\c c}on model, which closely matches that detailed in \citet{Awiphan2015}. From a sample of $1691$ simulated {\it Roman} microlensing events with detectable planets from \citet{Penny2019}, the brightest $432$ (26\%) microlensing events with lenses brighter than $W149=22.5$ were added to the images; the $W149$ and $VIS$ magnitude distributions of these samples are shown in Figure~\ref{figure:ImgSimInput}. The full sample of events\footnote{Available at \url{https://github.com/mtpenny/wfirst-ml-figures}} is a representative draw from the fiducial {\sc Roman} simulations, with stellar properties (e.g., proper motion, magnitude, etc.) drawn from an earlier version of the Besan{\c c}on model, BGM1106 as described in detail by \citet{Penny2019} and \citet{Penny2013}.The positions of all stars in the image were chosen randomly at a time $t=0$ (corresponding to the Roman launch date), and their positions at other times determined by their proper motion. Source stars were placed randomly in the image, but lens stars were placed such that they would recreate the parameters of the simulated microlensing events given the relative proper motion of source and lens. {\sc Euclid} VIS magnitudes were assumed to be equal to $Z087$ magnitudes from {\sc Roman}, and the simulated sources and lenses magnitudes were calculated by integrating over an interpolation of the stellar spectral energy distribution output by the Besancon model in the $R$, $I$, $Z$, $J$, and $H$ bands, plus the magnitudes of a blackbody with the same temperature, radius, distance, and extinction for the $K$ and $L$ bands. 

We set the {\sc Euclid} observations at $t=-1800$~days (i.e. about 5 years prior to the {\sc Roman} launch), and generated a set of 16 $1.83'\times 1.83'$ images at the native resolution of the VIS instrument. Each of the 16 images were dithered by a set of randomly chosen offsets between 0-10 pixels in the $X$ and $Y$ directions. The 16 images at each epoch were stacked using the drizzle algorithm~\citep{Fruchter2002} to refine the resolution to 0.0275 arcsec/pixel.

\subsection{Modeling process and results}
We first estimated an empirical and flux-calibrated point-spread function (PSF) of the drizzled images by modeling isolated stars, and we assumed a constant PSF. In practice, the PSF will vary across the field but will be extremely well-characterized. Assuming that the lenses and sources are resolved when the separation is $\gtrsim 0.25$ FWHM (i.e. 1.8 pixels in this case), 64 \% of events are resolved at the time of the simulated Euclid observation. However, with a median separation of 2.1 pixels, a lot of events are just at the limit of separation. We rejected three events from the analysis because they were located too close to the images edges. We then fit the 429 remaining simulated events, assuming that the light present in small stamps around the event location (estimated for the event coordinates at $t_0$) is solely due to the lens, the source, and a constant background The fit parameters are the total flux $f_t = f_s+f_l$, where $f_s$ and $f_l$ are the source and lens fluxes, the flux ratio $q=f_s/f_t$, the proper motion of the source and the lens, and the background level. We used the first image from Euclid to estimate the potential of early Euclid observations.

\begin{figure*}[h]
    \centering
    \includegraphics[width=.95\textwidth]{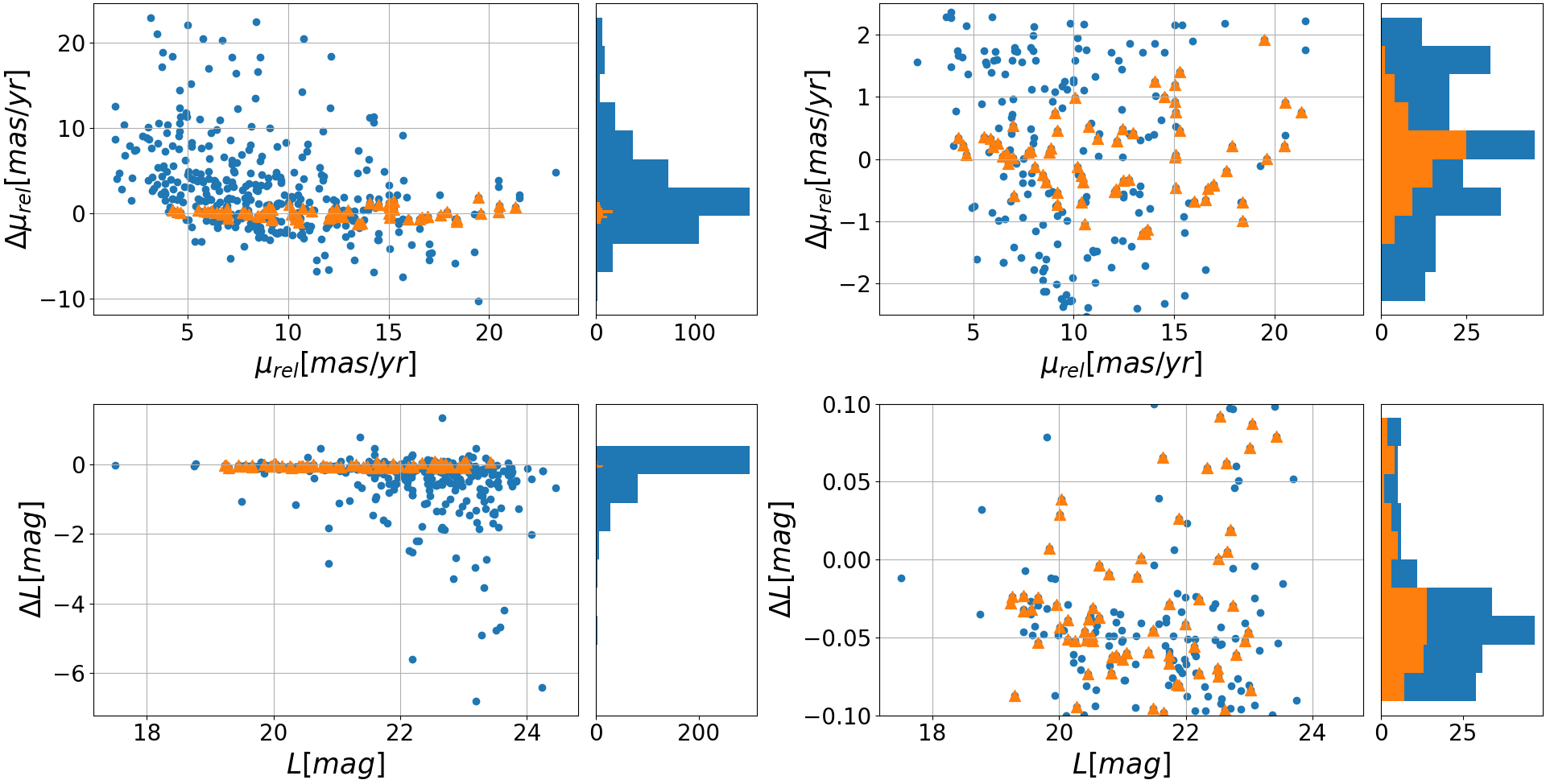}
    \caption{Distribution of fit residuals, defined as $\Delta = \rm{fit-true}$, as a function of $\mu_{rel}$ (top) and the VIS lens magnitude L (bottom). Orange triangles are events meeting both criteria: $|\Delta \mu_{rel}|<0.1\mu_{rel}$ and $|\Delta L|<0.1 $ mag. The left column represents the results for all of the 432 events, while the right column is a zoom for events with $|\Delta \mu_{rel}|<2.5$ mas/yr and $|\Delta L|<0.1$ mag.}
    \label{figure:EuclidFits}
\end{figure*}

The results of the fits are presented in the Figure~\ref{figure:EuclidFits}. We find $\mu_{rel}$ to be constrained, at better than the 10\% level (relative), for 29\% of the fits. We also find that the magnitude of the lens is reconstructed at better than 0.1 mag for 42\% of our sample. 15\% of the events are reconstructed within the two preceding constraints, and 85 \% of them are resolved (with a median separation of 3.5 pixels). For this subset of events, the properties of the lens will be reconstructed with high fidelity according to Eq.~\ref{eq:HR} and Eq.~\ref{eq:te}. Indeed, while the mass and distance degeneracy can persist for the lightest lenses \citep{Yee2015}, \citet{Bachelet2019} demonstrates that the parallax will be constrained for most events. Scaled to the 30,000 events expected from the {\sc Roman} mission, this indicates that early {\sc Euclid} observations can constrain the properties of over 2000 lenses, increasing our knowledge of the microlens distribution by several orders of magnitudes.

The analysis presented is relatively CPU-intensive, and this was the primary motivation to limit the sample to lenses with $W149<22.5$ mag. While this sample is not exactly representative, we note that the distribution of the source magnitudes is much broader (with $W149<26$ mag). We did not find any significant trends in the fit results that depends on the source magnitudes, indicating that our fits would also be reliable for fainter lens, since the problem is symmetric. We note however that our fitting approach, while generally accurate, is sub-optimal for a significant fraction of the events. This is mainly due to the presence of unrelated stars near to the line of sight, but this problem can be tackled by a more in depth analysis such as that done by \citet{Bhattacharya2017}. This can also result from a large difference in the lens and source brightness, coupled with a small lens/source proper motion. A complete understanding of the fitting performance (for example the accuracy of the PRF and the noise models of drizzle images) is beyond the scope of this study, as our goal here is to highlight the potential of early {\sc Euclid} observation. However, by studying each event individually, as would be the case with real data, it is likely that the performance could be significantly improved. The results presented here can therefore be considered conservative.
 
To conclude, a single {\sc Euclid} observation prior to the {\sc Roman} mission will place strong constraints on the mass and distance for thousands of lenses for a relatively small observational cost. Indeed, assuming that the microlensing {\sc Roman} fields will be of 2 $\rm{deg}^2$ \citep{Penny2019}, it would take about four pointings of {\sc Euclid} to cover the region of interest. Using the observing strategy presented in the previous section (16 dither images, each with a 300 s exposure time), this represents 7 hours of {\sc Euclid} telescope time with overheads. This strategy, coupled with the parallax measurements from the {\sc Roman} light curve (see Section~\ref{sec:para} and \citet{Bachelet2019}), will provide unprecedented constraints on masses and distances of thousands of lenses, ultimately placing exquisite constraints on the Galactic demographics of planets \citep{CalchiNovati2015}. It will also permit a direct measurement of the masses and distances of the earliest {\sc Roman} microlensing events, without the need to await the end of the {\sc Roman} mission. 

\section{Simulating joint {\sc Roman-Euclid} observations of free-floating planets}
\label{simul}
\subsection{Examples of joint observations}

Planetary-mass lenses (assumed here as $M\le 13 ~M_{\rm Jup}$) have extremely small Einstein radii and therefore microlensing events due to such lenses can display strong finite-source effects (i.e., $\rho_*\geq u_0$ from Eq~\ref{star-size}), allowing the measurement of $\rho_*$, and therefore $\theta_{\rm E}$. Obtaining parallax parameters ($\pi_{\rm E}$ or $\pi_{\rm rel}$) for FFPs therefore means that the FFP mass and distance can be directly measured. It is worth emphasising that such microlensing measurements would yield not just the FFP mass but also most of the FFP's full phase space (with the exception of velocity along the line of sight). This may provide vital information on the mode of FFP formation. 

\begin{figure*}
\centering
\includegraphics[width=.9\textwidth]{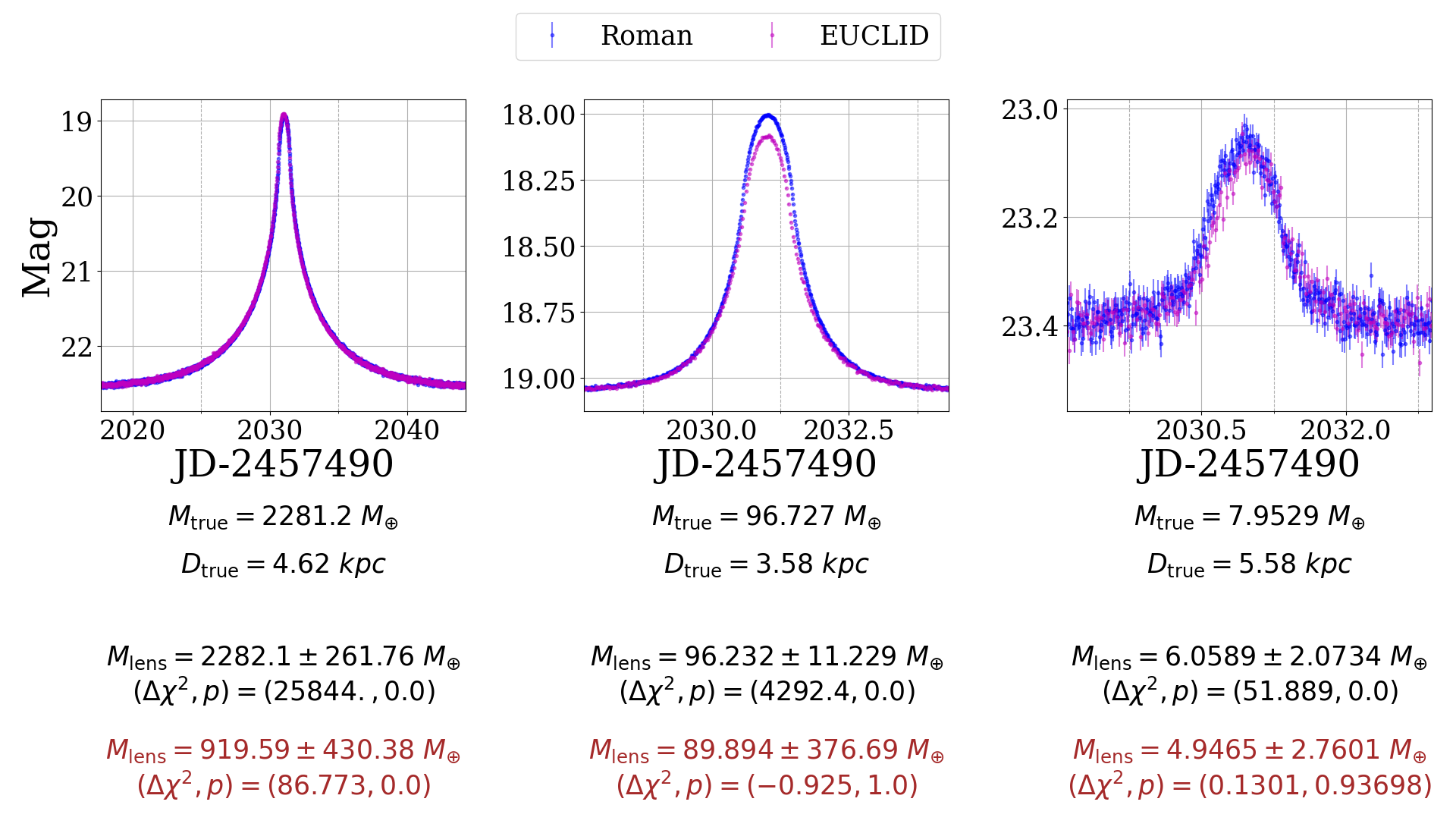}
\caption{\ Three examples examples of simulated microlensing events due to FFP lenses as seen by {\sc Euclid} and {\sc Roman}. From left to right are represented the case of a super-Jupiter lens, a Saturn-like lens and a super-Earth lens. The distance of the source $D_S$ is set at 8 kpc and the true mass $M_{true}$ and distance $D_{true}$ of the lenses are indicated for each cases, as well as the best-fit parameters. Those based only on {\sc Roman} data are presented in brown, while shown in black are best-fit solutions using both datasets. Using {\sc Roman} data only, the mass of the lens can be reconstruct accurately only for the longer events. Magnitudes are artificially aligned to the {\sc Roman} system for the plotting. }
\label{fig:example_lc}
\end{figure*}

To illustrate such measurements, we simulate three examples of simultaneous observations made by {\sc Roman} W149 and {\sc Euclid} NISP (H), assuming a phase separation of $\sim \pi/4$ between the orbits of the telescopes around the Earth-Sun L2. The photometric precision has been consider similar for both telescopes, while the cadence of {\sc Euclid} and {\sc Roman} are set to 30 min and 15 min, respectively. We use the pyLIMA software to simulate and fit the lightcurve \citep{Bachelet2017}, and use the Gaia ephemerides around the Earth-Sun L2 to obtain realistic orbits. For each cases, we model the event with and without including the parallax effect. We derive the log-likelihood ratio $\Delta \chi^2$ associated with a p-value $p$, indicating if the fit including parallax is statistically more significant than the simpler model. We run two set of fits: one including both telescopes and one using only {\sc Roman} data. Results are presented in Figure~\ref{fig:example_lc}. We derived the lens masses and errors assuming $\theta_{\rm E}$ to be known to within 10\%. This implies $\rho$ to be measured, and therefore we also force $u_0\le\rho$. We assume a linear limb-darkening law and using $\Gamma=0.5$ \citep{Yooo2004}. 

\subsection{Simulation of joint observations}

In this section, we explore, in a statistical sense, the capabilities that offer a {\sc Roman} and {\sc Euclid} joint-survey to detect and characterize microlensing events due to FFPs.  Therefore, this work is in the lineage of the previous work from \citet{Johnson2020} and \citet{Ban2020}, but, as detailed below, implements several refinements that can play a significant role on the characterization (i.e. the mass measurement) of events. To list a few, this includes the consideration of limb-darkening for a more accurate magnification estimation, an updated version of the Besan\c{c}on model calibrated to \textsc{Hubble Space Telescope} and compared to the OGLE-IV observations and the use of time-integrated selection cuts, that are more reliable to ensure the ability to measure finite-source effects and parallax, to ultimately better characterize events.

To assess the number of FFPs where such mass constraints could be obtained, we used a specially modified version of the \mabuls{} simulator\footnote{\url{www.mabuls.net}} \citep{Specht20} to allow for FFPs and to consider simultaneous parallax by separated observatories. \mabuls{} employs a synthetic population of microlensing lenses and source stars seeded from version 1307 of the Besan\c{c}on Galactic model (BGM) \citep{Robin2012,Robin2014}. \mabuls{} has been demonstrated by \citet{Specht20} to provide an accurate match to the 8,000-event sample from the OGLE-IV survey \citep{Mroz19b}. Using the BGM, \mabuls{} simulates the microlensing event rate from the synthetic stellar catalogues produced.

The BGM divides the Galaxy into four components; the thin disk, bulge \citep{Robin2012}, thick disk and halo \citep{Robin2014}, each with their own stellar initial mass functions, density laws and kinematics. It also includes a 3D extinction model from \citet{Marshall06}. FFPs of various mass are injected to replace stars as lenses. The FFP mass functions that were considered included two Dirac delta functions, one peaked at Earth mass with a normalisation of 10 FFPs per main sequence star, as suggested by \citet{Mroz2019} and another peaked at Jupiter mass with a normalisation of 2 FFPs per main sequence star, as suggested by \citet{Sumi2011}.

Finally, FFPs inherit the kinematics of the stars they replace in the synthetic catalogue. The overall procedure for computing the microlensing rate, optical depth and average timescale follows the formalism detailed in \cite{Specht20} with a few differences. Firstly, our signal-to-noise (S/N) selection involves a time-averaged statistic rather than the S/N at peak used in \cite{Specht20}. Specifically,  we demand a $\Delta\chi^2$ between the synthetically-generated microlens lightcurve and a best fit constant flux model of at least 125, with at least 6 points reaching $3\sigma$ above the baseline. 

In addition, the joint detection of space-based parallax between the {\sc Roman} and {\sc Euclid} light curves is required to be at least $5\sigma$, following the Fisher matrix analysis of \citet{Bachelet2019}, with modifications to account for the BGM formalism. A detection of finite source effects is also demanded, with a minimum $\Delta\chi^2$ between a PSPL and finite source model of 100. We apply a maximum impact parameter threshold value $u_{\rm t} = 3$ for $u_{\max}$, corresponding to a minimum required peak magnification of 1.017, following the formalism for $u_{\rm max}$ from \citet{Specht20}. The properties of each filter used in this simulation are shown in Table~\ref{table:table0}.

\begin{table}[]
\caption{Assumed sensitivities for the {\sc Roman} W146 and {\sc Euclid} VIS and NISP-$H$ band passes used in our simulations. Tabulated are the point spread function solid angle $\Omega_{\rm psf}$, the zero-point magnitude of the filter $m_{\rm zp}$, the exposure time $t_{\rm exp}$, the sky background $\mu_{\rm sky}$, the observation frequency and the duration of the Galactic observing season. The two {\sc Euclid} bulge observing campaigns per season are each assumed to be fully contained within the respective {\sc Roman} campaigns.}
\label{table:table0}
\centering
\begin{tabular}{cccc} 
    \hline\hline
     & \textbf{W146} & \textbf{VIS} & \textbf{NISP-$H$}\\ 
    \hline
    $\Omega_{\rm psf}$ ($\rm arcsec^2$) & 0.0456 & 0.0254 & 0.1590 \\ 
    $m_{\rm zp}$ & 27.62 & 25.58 & 24.92 \\
    $t_{\rm exp}$ ($\rm sec$) & 46.8 & 270 & 54 \\
    $\mu_{\rm sky}$ ($\rm mag$ $\rm arcsec^{-2}$) & 21.5 & 21.5 & 21.4 \\
    Cadence ($\rm min$) & 15 & 60 & 60 \\
    Season (days) & $2\times72$ & $2\times 30$ & $2\times 30$ \\
    \hline
 \hline
\end{tabular}
\end{table}

The value of the maximum detectable impact parameter, $u_{\max}$, for each simulated event is obtained separately for the S/N criterion ($u_{S/N}$), the parallax criterion ($u_{\rm plx}$) and the finite source criterion ($u_{\rm FS}$), with the final value taken as $u_{\rm max} = \min(u_{S/N},u_{\rm plx},u_{\rm FS})$. In each case, multi-dimensional linear interpolation is performed on a pre-computed $u_{\max}$ look-up table, to reduce computation time and prevent the necessity of performing expensive finite source calculations at runtime. The S/N look-up table was three-dimensional, with $u_{\rm max}$ depending on the source magnitude, the cadence normalised to the event timescale, $x$, and the normalised source radius $\rho_*$. The parallax look-up table was six-dimensional, depending on the source magnitudes in the {\sc Roman} W146 filter $m_{\rm W}$ and the {\sc Euclid} filter (either the VIS $RIz$ filter or the NISP $H$-band filter - $m_{\rm E}$), $t_{\rm E}$, $\rho_*$, $\pi_{\rm E}$ and the angle between the projected baseline vector and the $\vec{\mu_{\rm rel}}$ vector, $\phi$. Finally, the finite source look-up table was also three-dimensional, depending on the same input parameters as the S/N table. The parameters and grid resolutions for each of the three $u_{\rm max}$ grids are shown in Table \ref{tab:grid_params}.

\begin{table}[]
    \caption{The input parameters for each of the $u_{\rm max}$ grids are shown, with the parameter ranges shown in square brackets. Parameters labelled with $\dagger$ are distributed logarithmically, with all others distributed linearly. All parameters have a grid resolution of 10 points, other than $\phi$ and $\rho_*$ for the parallax grid which have a grid resolution of 5 points. The parameters used throughout are $m_{\rm S}$ an arbitrary source magnitude, $m_{\rm W}$ a W146 magnitude, $m_{\rm E}$ a magnitude in either {\sc Euclid} VIS or NISP(H) filters, $t_{\rm E}$ the Einstein radius crossing time, $x$ the cadence of a telescope normalised to $t_{\rm E}$, $\rho_*$ the source radius normalised to $\theta_{\rm E}$, $\pi_{\rm E}$ the magnitude of the microlensing parallax and $\phi$ the angle between the lens-source proper motion vector and the projected baseline vector.}
    \centering
    \resizebox{9cm}{!}{\begin{tabular}{cc}
    \hline\hline
    \textbf{$u_{\rm max}$ grid type} & \textbf{Input parameters} \\
    \hline
        Signal-to-noise & $m_{\rm s}$ [16,24.9], $x$ [0.01,1.0]$^\dagger$, $\rho_*$ [0,5] \\
        Parallax & $m_{\rm W}$ [16,24.9], $m_{\rm E}$ [16,24.9], $\pi_{\rm E}$ [$10^{-4},1$]$^\dagger$, $t_{\rm E}$ [0.02,6], $\phi$ [0,$\frac{\pi}{2}$],  $\rho_*$ [0,5] \\
        Finite Source & $m_{\rm s}$ [16,24.9], $x$ [0.01,1]$^\dagger$, $\rho_*$ [$10^{-5}$,5]$^\dagger$ \\
        \hline
    \hline
    \end{tabular}}
    \label{tab:grid_params}
\end{table}

\begin{figure}
\centering
\includegraphics[width=9cm]{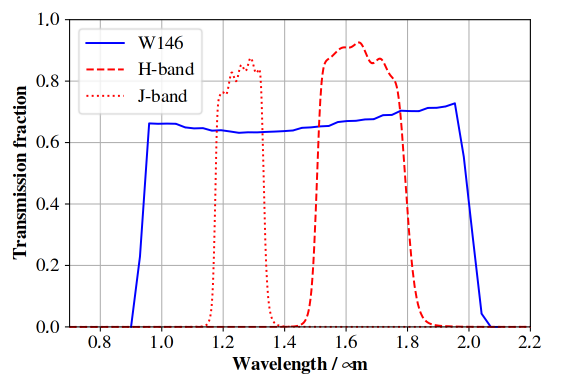}
\caption{The transmission curves for the {\sc Roman} W146 and Johnson-Cousins $J$ and $H$ filters are shown. Data for the W146 transmission is available at: \protect\url{https://wfirst.gsfc.nasa.gov/science/WFIRST_Reference_Information.html}}
\label{figure:W146}
\end{figure}
Calculating the {\sc Euclid} VIS and {\sc Roman} W146 magnitudes for each source star was achieved by using approximations based on Johnson-Cousins filters. For VIS we used the Johnson-Cousins $R$-band, which most closely mimics the VIS wavelength coverage and the wavelength of peak transmission. For W146, a weighted composite of the Johnson-Cousins $J$ and $H$ bands was used; shown in figure \ref{figure:W146} are the transmission curves for $J$, $H$ and W146 for comparison. The resulting W146 magnitude $m_{W146}$ is given by
\begin{equation}
    m_{w146} = -2.5log_{10}(10^{-0.4m_J} + 10^{-0.4m_H}) + \delta m,
\end{equation}
where $\delta m$ accounts for the difference in the transmission integrals of W146 and the combination of $J$ and $H$.

\begin{figure*}[h]
    \centering
    \includegraphics[width=.9\textwidth,height=.7\textheight]{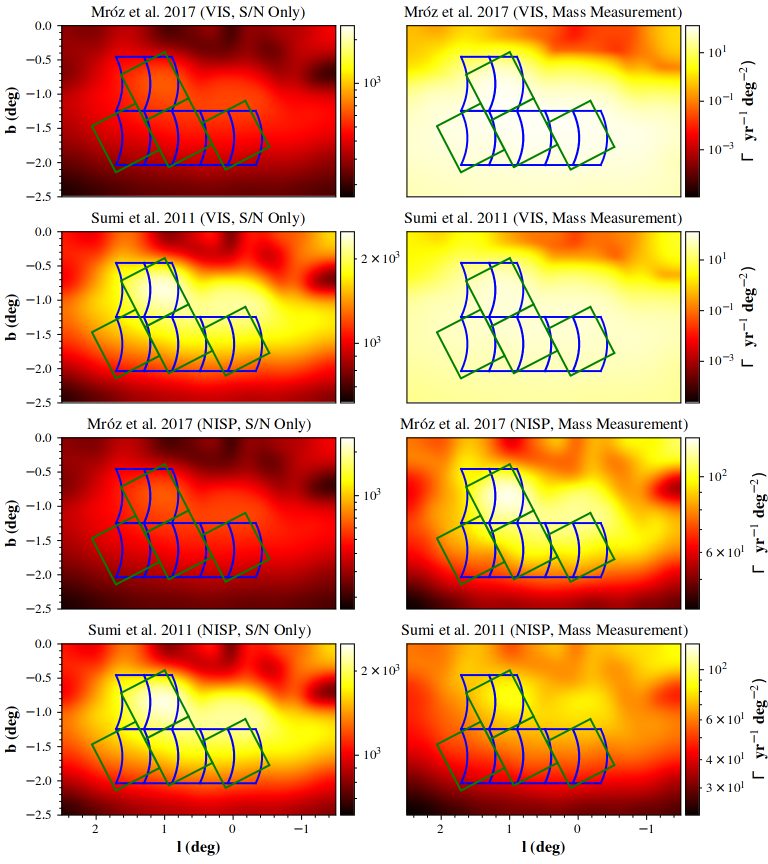}
    \caption{FFP rate maps, in units of events per square degree per year, for events jointly detected by both the {\sc Roman} W146 filter and {\sc Euclid} VIS (top two rows) and NISP (bottom two rows) filters. Rows one and three use the Mr{\'o}z model of ten Earth mass FFPs per main sequence star, while rows two and four use the Sumi model of two Jupiter mass FFPs per main sequence star. The left column shows rate maps constrained only by the signal-to-noise criterion, while the right column shows the equivalent rate maps with all selection criteria present, including parallax and finite source effects.}
    \label{figure:FFP_MAPS}
\end{figure*}

\begin{figure*}[h]
    \centering
    \includegraphics[width=.9\textwidth,height=.7\textheight]{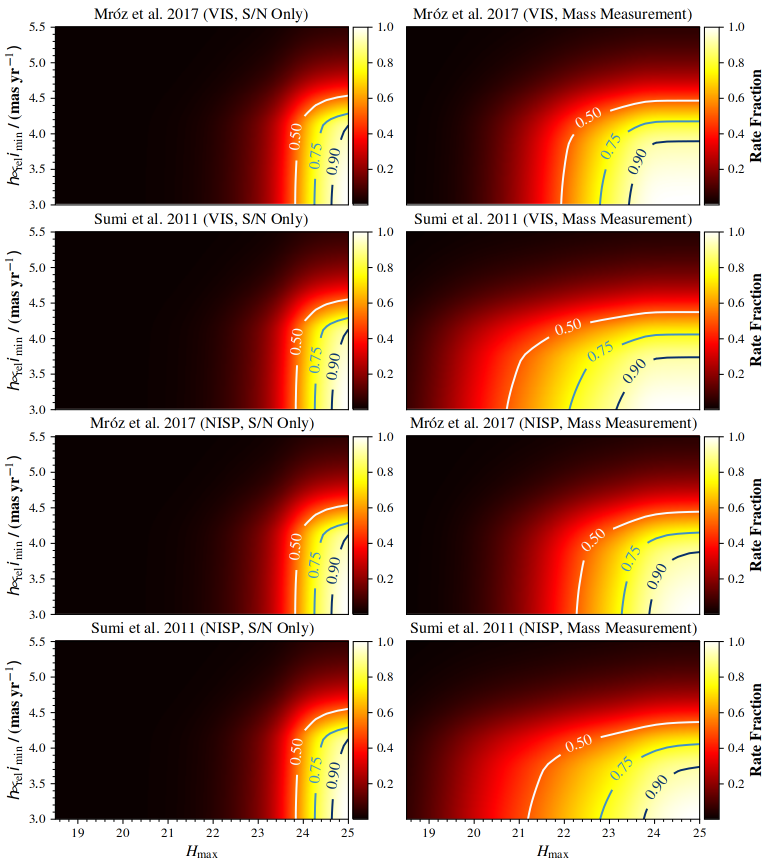}
    \caption{The cumulative rate fraction for a maximum \textit{H}-band magnitude and minimum $\langle \mu_{\rm rel} \rangle$ is shown in both the {\sc Roman} W146 filter and {\sc Euclid} VIS (top two rows) and NISP (bottom two rows) filters. Rows one and three use the Mr{\'o}z model of ten Earth mass FFPs per main sequence star, while rows two and four use the Sumi model of two Jupiter mass FFPs per main sequence star. The left column shows rate fraction constrained only by the signal-to-noise criterion, while the right column shows the equivalent rate fraction with all selection criteria present, including parallax and finite source effects. The 0.5, 0.75 and 0.9 contours are shown for reference.}
    \label{figure:FFP_FRAC}
\end{figure*}

The results of the simulation, which used approximately $1.5\times 10^{11}$ unique lens-source pairs, were compiled into microlensing event rate maps over Galactic coordinates $l \leq 2.5^{\circ}$, $l \geq 358.5^{\circ}$ and $-2.5^{\circ} \leq b \leq 0^{\circ}$, which covers the {\sc Roman} Cycle-7 field locations proposed by \citet{Penny2019}. Four {\sc Euclid} fields were then added with dimensions $0.76^\circ \times 0.72^\circ$ \citep{Penny2013}. Since the relative field rotation between the two is as yet unknown, an "unoptimised" scenario of field alignments was considered, where the Cycle-7 fields were aligned with the Galactic coordinate system, while the {\sc Euclid} fields were aligned with the ecliptic coordinate system. Hence the numbers provided in this work for total rates are deemed conservative. The {\sc Euclid} field placements were optimised to maximise the joint {\sc Roman-Euclid} detection rate in the {\sc Euclid} NISP (\textit{H}) filter for the {Mr\'oz} model, using Nelder-Mead maximisation that samples the rate maps only in the overlap region between the two surveys. These field locations were then used for all other rate maps, with results shown in figure \ref{figure:FFP_MAPS}. An extra condition was applied requiring that the {\sc Euclid} fields be contiguous and non-overlapping. To show illustrate the fraction of events with $\mu_{\rm rel} > \langle \mu_{\rm rel} \rangle_{\rm min}$ and for source stars brighter than $H < H_{\rm max}$, complimentary cumulative rate fraction plots were generated, shown in figure \ref{figure:FFP_FRAC}.

Predicted joint detection rates are displayed in Table~\ref{table:table1}. The numbers indicate that a joint survey would have parallax sensitivity down to Earth mass FFPs. In our simulation, the parallax was measurable for 91\% of events with a Jupiter mass lens and 85\% for events due to a Earth lens. Therefore, almost all FFP events observed by the joint Euclid-Roman survey will have at least one mass-distance relation constrained. The detection of finite-source effects occurs for 4\% of Jupiter FFPs events and 21\% of Earth events. For this smaller fraction, the mass and distance of the FFP will be known with high precision.

A comparison can be made to the parallax detection rates from \citet{Ban2020}, who used different methods for determining the event selection criteria along with microlensing calculations based on an earlier version of MaB$\mu$lS. \cite{Ban2020} considers a single line of sight at $(l,b) = (1.0^{\circ},-1.75^{\circ})$ for various telescope combinations (including the Vera Rubin Observatory, formerly LSST). 
Differences in calculation methodology make direct comparison difficult. \cite{Ban2020} does not require a detection of finite source effects (although finite source effects without limb-darkening are taken into consideration when calculating simulated lightcurve photometry) and uses a signal-to-noise at peak selection (with S/N $\geq$ 50), as opposed to the time-integrated signal to noise used in the present study. This difference in signal-to-noise selection drives the main difference between results, with a rate per square degree per 60 days of observation of $\Gamma_{\rm Earth} = 5.2$ for this work and $\Gamma_{\rm Earth} = 4.9$ for \citet{Ban2020}, with a larger discrepancy for the rates of Jupiter mass FFPs of $\Gamma_{\rm Jupiter} = 104$ for this work and $\Gamma_{\rm Jupiter} = 31$ for \citet{Ban2020}, attributable to the effect of using a time integrated signal-to-noise selection on events of longer timescales. Similarly, we can compare the fraction of FFP events with finite-source effects found in this study with the fraction estimated in \citet{Johnson2020}. The criterion for the detection of finite-source used in this work ($\delta_\chi^2\ge100$ between a FSPL and a PSPL model) can be approximated as $\rho\gtrsim u_0$. Reading the fraction of detected events with $\rho>u_0$ on the Figure 9 of \citet{Johnson2020} returns few \% for Jupiter-mass FPPs and $\sim$ 15 \% for the Super-Earth case (i.e. $10 M_\oplus$), in good agreement with the estimated fraction presented in this work (4\% and 21\% respectively). We note an even better agreement if one choose the  same criterion as \citet{Johnson2020} for detectable finite-source effects, i.e. $\rho\ge0.5u_0$.

It is clear from Table~\ref{table:table1} that even a single-season {\sc Roman-Euclid} joint campaign has the potential to detect and verify the existence of FFPs down to Earth mass, or begin to place strong limits on their abundance, limits more than five times stronger than current limits \citep{Mroz2017}. With a multi-season joint campaign, the sensitivity increases proportionately. If FFPs have an abundance comparable to one per Galactic star, then a joint campaign can obtain direct mass, distance, and kinematic measurements for a significant sample, providing a high-precision test of FFP formation models.

\begin{table*}[]
\caption{Predicted joint {\sc Roman-Euclid} FFP microlensing detection rates per annual observing season, consisting of two 30-day Galactic bulge observing windows per season. Each of the 30-day {\sc Euclid} windows occur within a corresponding 72-day observing window for {\sc Roman}. The signal-to-noise and parallax criteria required for joint detection are discussed in the main text. Columns 3-6 show the effect of introducing different combinations of event selection criteria on the detection rate. Where  no parallax or finite source measurement is possible (``S/N only'') only a statistical order-of-magnitude FFP mass measurement is possible. Where either parallax or finite source size is measured the three-way microlens degeneracy is partially broken resulting in much improved statistical mass determinations (around a factor 2 uncertainty). When both parallax and finite source are measured (``All constraints'') the microlens degeneracy is fully lifted and a direct mass measurement is possible. The ability of Roman and Euclid to work together to measure parallax allows a huge improvement in the fraction of events ($85-90\%$) where the parameter degeneracy is partly or fully broken.}
\label{table:table1}
\centering
\begin{tabular}{cccccc} 
    \hline\hline
    \textbf{FFP Model} & \textbf{Filter Combination} & \textbf{S/N only} & \textbf{S/N + Parallax} & \textbf{S/N + Finite Source} & \textbf{All Constraints} \\ 
    \hline
    \multirow{2}{*}{Sumi (2011)} & W146 + VIS & $490$ & $450$ & $18$ & $18$ \\ 
    & W146 + NISP (\textit{H}) & $490$ & $450$ & $19$ & $19$ \\ 
    \hline
    \multirow{2}{*}{Mr{\'o}z (2019)} & W146 + VIS & $130$ & $110$ & $28$ & $28$ \\ 
    & W146 + NISP (\textit{H}) & $130$ & $110$ & $31$ & $31$ \\ 
    \hline
 \hline
\end{tabular}
\end{table*}

\section{Exomoons} 
\label{sec:exomoons}

\begin{figure*}
\centering
\includegraphics[width=.9\textwidth]{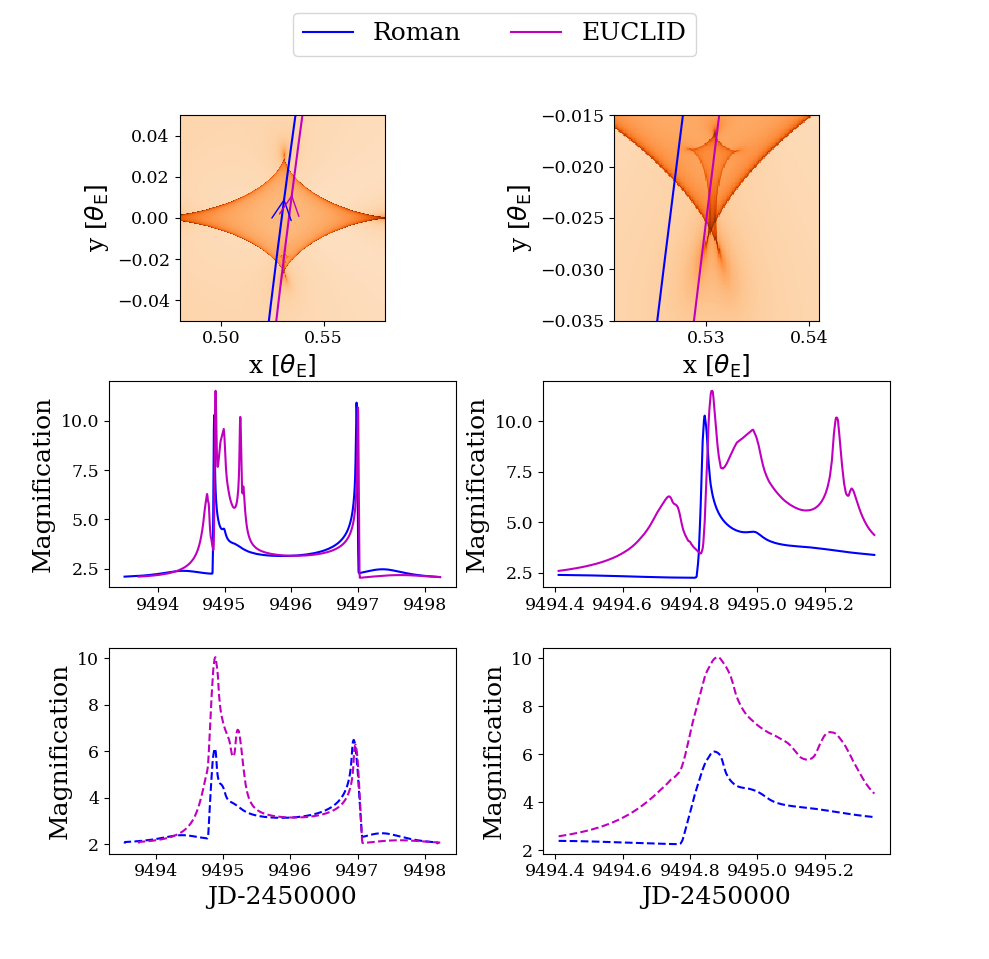}
\caption{Simulated example of an exomoon event---a triple-lens caustic----observed from both {\sc Roman} and {\sc Euclid}. The top row displays the event geometry centred on the planetary caustic. The right column is a zoom-in on the time of the exomoon caustic crossing. The expected separation of the source tracks is comparable in scale to the perturbation induced on the caustic by the exomoon. The planet, with mass ratio $q_{1}=10^{-3}$, is at separation $s_{1}=1.3$, and the moon has $q_{2}=10^{-2}$ and $s_{2}=0.032$ relative to the planet. The Einstein timescale is $t_{\rm E}=47~{\rm days}$. The middle row is for a normalised source radius $\rho=0.00033$ while the bottom row is for $\rho=0.001$.}
\label{figure:exomoons}
\end{figure*}

The approach of combining {\sc Roman} and {\sc Euclid} observations not only provides the opportunity for detecting planetary events of lunar mass \citep{Penny2019}, but would in tandem increase the chances of discovering extrasolar moons. Microlensing detections of exomoons has been suggested by \cite{han2002} and simulated by \cite{liebig2010}. Exomoons are effectively described by triple point-mass lens models. As such, the number of caustic curve topologies and the respective variety of light curves increases substantially, compared to binary lenses \citep{danek2015}. In most cases, one can expect that the mass ratio of the lunar companion with respect to the host star is below $10^{-7}$ (the mass ratio of Ganymede relative to the Sun is $q\sim7.5\times10^{-8}$) and thus will lead to a perturbed magnification pattern lasting a few hours at maximum. Although earlier works have emphasized detecting exomoons, from the source tracks shown in Fig.~\ref{figure:exomoons}, it is clear that the separation of the tracks is comparable to the perturbation induced in the caustic curve, which could double the probability of detecting exomoons in an optimistic scenario. 

To simulate the two source trajectories, we assumed that both {\sc Euclid} and {\sc Roman} will have orbital elements similar to those of {\sc Gaia}, and therefore we used the {\sc Gaia} ephemeris to simulate the observations. Ideally, the detection of exomoons would require an armada of space telescopes simultaneously probing planetary caustics, well beyond the scope of the present concept. Even then, characterizing an exomoon would remain challenging, and securing a second light curve covering the exo-lunar caustic perturbation is essential to place a constraint on the exo-lunar mass ratio and separation from the planet.

As we are dealing with an uncharted region of the observable exo-lunar mass and semi-major axes parameter space, it is hard to estimate the number of detections. Given the frequency with which moons occur around Solar System planets\footnote{There are, for instance, 212 satellites on NASA's list of planetary satellites \textit{https://ssd.jpl.nasa.gov/sats/elem/}}, it is highly likely that at least some of the $\sim 1500$ cool-orbit exoplanets that {\sc Roman} will discover will host exomoons and that by combining data from {\sc Roman} and {\sc Euclid}, we will be able to confirm the object's status as exomoons and establish in a statistically significant sense if exomoons are common.

\begin{figure*}[!ht]
\centering
\includegraphics[width=.9\textwidth]{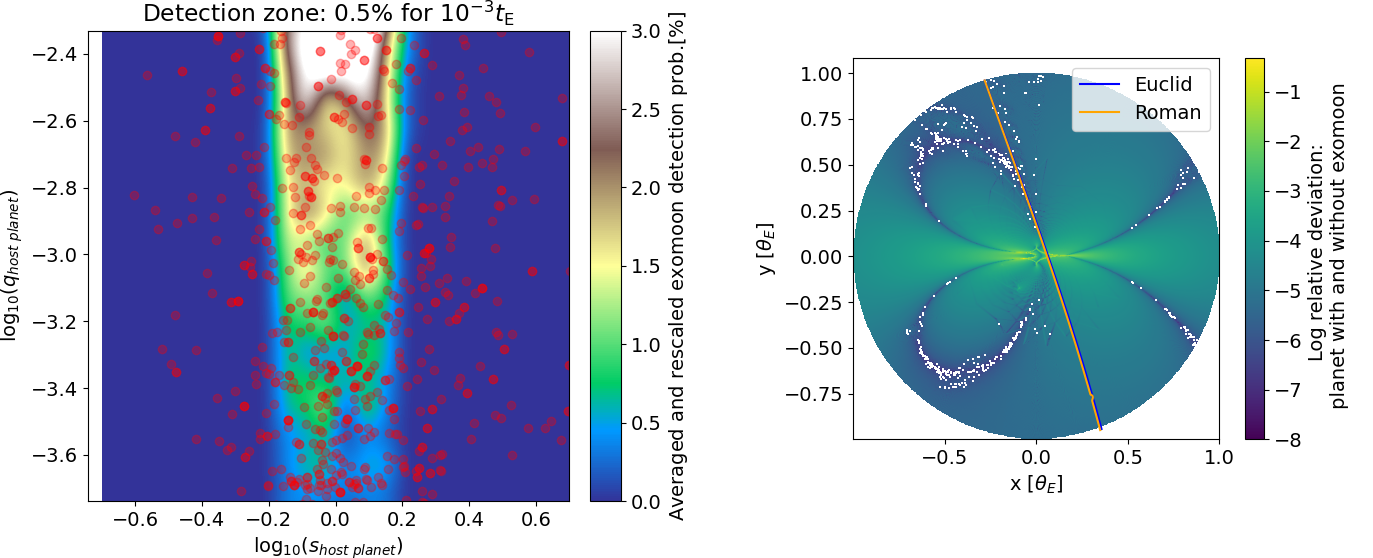}
\caption{The simulated detection zone areas were converted to assess probability for exomoons around host stars (shown as red dots) assuming a detection threshold of 0.5\,\% which would correspond to 5\,mmag accuracy for unblended events - well within the expectation of the {\sc Roman} telescopes. In order to account for the sampling interval we require consecutive points to be within $10^{-3} t_{\rm E}$ to be detectable. For more massive planets (green) we expect the detection zone to cover $10^{-3}\,\theta_{\rm E}^2$. Just for reference, the Einstein radius of an isolated exomoon with $q_2 = 10^{-5}$ would be roughly $10^{-3}$ smaller than the Einstein radius of the host star. A representative sample of planets detected by {\sc Roman} \citep{Penny2019} is used in order to asses the number of detectable exomoons. In addition, the logarithmic relative deviation and simulated tracks are shown for both missions for illustrative purposes.}
\label{figure:exomoon_zone}
\end{figure*}

In order to put the exomoon detection probability into context, one can introduce an exomoon lensing signature into the binary models of the full sample of microlensing exoplanets discovered to date. Little is known about the distribution of exomoons but based on the so far known microlensing exoplanets and deriving exomoon properties (mass and orbital radius) based on the distribution of moons in the Solar System, we can estimate the detection zone on a simulated magnification pattern under the following assumptions: the ratio of the exomoon mass to the host star mass ratio $q_2 \in \left[10^{-5}, 10^{-2}\right]$ is log-uniform distributed. The angular separation  $s_2$ between host planet and exomoon similarly was log-uniform distributed and is expressed in units of $\theta_{\rm E}$  for a range of $s_2 \in \left[10^{-3}, 10^{-1}\right]$. In addition, each exomoon should be uniformly distributed on a sphere around the exoplanet. The detection zone is shown in Fig.~\ref{figure:exomoon_zone} and for randomly oriented tracks and lensed events, i.e. within $1~\theta_{\rm E}$, a detection zone area of $10^{-3}\,\theta_{\rm E}^2$ would correspond to a detection probability of $\approx 3\,\%$. 

In order to corroborate such a heuristic approach and to reach a more conservative estimate, one can simulate the parallax of microlensing events based on the range of lensing parameters for all events documented in the NASA exoplanet archive and analyze those parts of the lightcurve with a magnification $\mu>1.34$ - representing the lensing zone. For a planet in the lensing zone with $q_1 = 10^{-3}, s_1 = 1$ we find that for simulated exomoons and simulated tracks we obtain 0.9\% detectable events for a detection threshold of 0.5\% photometric accuracy which is in agreement with the integrated detection zone shown before. Within this work, both missions are supposed to have a similar sampling rate closer to the {\sc Roman} observing strategy. This leads us to the estimate that more than 0.8\% of exomoons are detectable for host planet parameter ranges in the control region of $\log_{10}(s_1) \in \left[-0.15,0.15\right]$ and $\log_{10}(q_1) \in \left[-3,-2\right]$. Less than 10\% of all expected extrasolar planets will be in that range and thus our final assessment will include a representative sample of the $s_1, q_1$ parameter space shown in Fig. \ref{figure:exomoon_zone}. Most events are covered by both missions and with a similar baseline of timeseries photometry, in which case, the number of detected exomoons would modestly increase. In total 40000 triple lens maps were simulated for the exomoon comparison with actual tracks. We find that 281 would be detected with both missions. The {\sc Roman} mission alone would contribute 25 more events and {\sc Euclid} (with a Roman-like cadence) would contribute another 25. That means the total number of exomoons will only decrease by less then 10\% if {\sc Roman} and {\sc Euclid} were not combined. The main impact is the improved characterization of individual exomoon events as indicated in Fig.~\ref{figure:exomoons}.

If the number of exoplanets is as high as predicted \citep{Penny2019} and exomoons are distributed in the aforementioned way, that could make the first exomoon detections feasible and certainly justifies a more detailed study of that subject. For that purpose, we are using the original detection zone map and convert that to a detection probability evaluated for a representative sample of \citet{Penny2019} for the host planet parameters $s$ and $q$. The respective would be 3 and thus on the order of one given the underlying assumptions.

\begin{figure*}
\centering
\includegraphics[width=.9\textwidth]{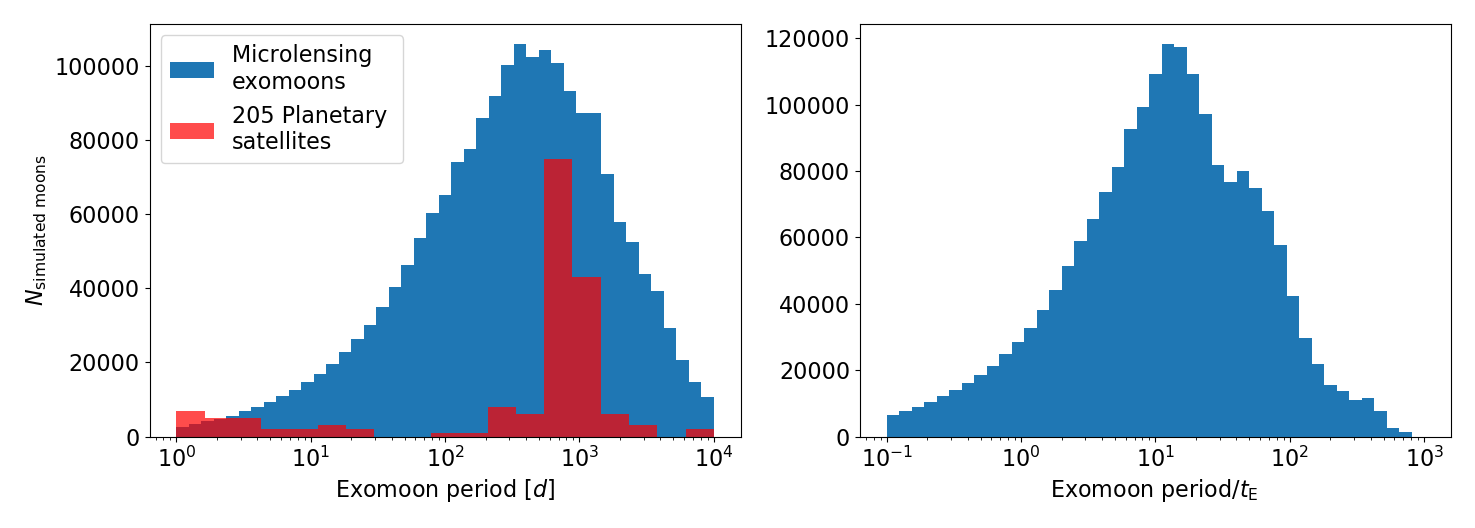}
\caption{Histogram of 2 million simulated exomoons for the default model of known microlensing exoplanets shown on NASA's exoplanet archive. For illustrative purposes a rescaled histogram of planetary satellites\protect\footnote{Based on NASA's planetary satellite mean elements \textit{https://ssd.jpl.nasa.gov/sats/elem/}} and their orbital periods are indicated. To assess the impact and the orbital motion the period is also shown with respect to the Einstein time.}
\label{figure:exomoon_periods}
\end{figure*}
Finally, we would like to highlight that the orbital motion was not simulated as second order effect. In the following, we are assessing the impact of the orbital motion by looking at a sample of simulated exomoons for the known default model of NASA's exoplanet archive. Since we are relying on distributions based on actual planetary detections, one can extract the orbital periods of the exomoons. That also enables us to express the period in terms of $t_{\rm E}$ as show  in Fig.~\ref{figure:exomoon_periods}. The mode of the distribution of periods is $\approx 10 t_{\rm E}$ which could contribute to the characterization and would lead to a detectable change in orbital motion.

\section{Summary}\label{sec:sum}
In this paper, we have assessed the benefits of {\sc Euclid} observations of {\sc Roman} microlensing fields. After recalling the different methods used to estimate the mass and distance of the lenses, we study the possibility of an early {\sc Euclid} imaging campaign (about 5 years prior to {\sc Roman} launch). We have simulated {\sc Euclid} and {\sc Roman} images of 432 microlensing events and modelled the proper motions and magnitudes of the sources and the lenses. Because of the larger separations between the sources and lenses at the time of the early {\sc Euclid} images, we were able to reconstruct 29\% of the simulated relative proper motions $\mu_{rel}$ at better than 10\%, using the {\sc Euclid} early image. Similarly, we have demonstrated that it is possible to reconstruct 42\% of lens magnitudes with a precision of 0.1 mag in the VIS band. We therefore conclude that the early imaging of the {\sc Roman} fields by {\sc Euclid} will allow a precise measurement of mass and distance for a large fraction of events that will be detected by {\sc Roman}, for a modest telescope time investment of about 7 hours, and this would be immediately available after the first year of {\sc Roman} observations.

We have further studied the potential of a joint simultaneous observing campaign, especially for constraining the microlensing parallax. We first show that this strategy allows the estimation of lens masses down to Earth-mass FFPs, a regime not achievable with {\sc Roman} alone. We simulated billions of microlensing events with two FFPs populations to estimate event rate maps towards the {\sc Roman} microlensing fields. Based on these maps and assuming three detection criterion, we found that hundred of events due to FFPs will be detected every year. Moreover, the combination of the two datasets will constraint the parallax for more than 80\% of events. For about 20\% of these events, finite-source effects will also be detectable and therefore the mass and distance of these objects will be known with high precision. We considered two different FFP population, constructed as Dirac delta functions, one peaked at Earth mass and normalised to 10 FFPs per main sequence star \citep{Mroz19b} and the second one peaked at Jupiter mass and normalised to 2 FFP per main sequence star \citep{Sumi2011}. Our results indicate that 490 Jupiter-mass FFPs and 130 Earh-Mass FFPs could be detected per year. \citet{Johnson2020} used different hypothesis, they considered the full observing season windows of Roman ($\sim$ 72 days) and one FFP per main sequence star in the galaxy, but their results are in good agreement with our estimation. Indeed, after correction of the hypothesis for the shorter Euclid observing seasons, the detection rate reported by \citet{Johnson2020} are 123 Earth-mass FFPs and about 550 Jupiter-mass FFPs per year. \citet{Ban2020} also assumed one FFP per main sequence star and studied three Dirac delta population peaking at Jupiter, Neptune, and Earth mass \citep{Ban2016}. By applying the same correction, their estimations scaled to with 152 Jupiter-mass and 123 Earh-Mass FFPs detected every year. We note that the discrepancy in the rate of Jupiter-mass FFPs is mostly attributable to the difference in the detection criteria between the two studies. Our results are therefore compatible with these previous studies and reinforce the claim that a joint Roman-Euclid survey will detect hundreds of FFPs, depending on the exact population of these objects. This an unique opportunity to study the FFP population in great details, especially to improve the picture on the abundance of FFPs, as well as place strong constraints on FFPs formation models.

Finally, we also study the potential of the joint-survey to detect exomoons. Using the distribution of moons in the Solar System and the planet distribution that {\sc Roman} is expected to detect, we simulated 40,000 triple lens events to estimate that about 1\% of these moons should be detectable, and we therefore conclude that this survey could lead to the first detection of an exomoon. In this scenario, the combination of the two datasets will be extremely valuable for the caracterization of the lensing system because the projected separation between {\sc Roman} and {\sc Euclid} is of the same order as the caustic size induce by the presence of the moon.


\begin{acknowledgements}
EB gratefully acknowledge support from NASA grant 80NSSC19K0291.The work of DS is funded by a UK Science and Technology Facilities Council (STFC) PhD studentship. EK also acknowledges support from the STFC. EB, JPB and CR's work was carried out within the framework of the ANR project COLD-WORLDS supported by the French National Agency for Research with the reference ANR-18-CE31-0002. JPB  was supported by the University of Tasmania through the UTAS Foundation, ARC grant DP200101909 and the endowed Warren Chair in Astronomy. JR was supported by NASA ROSES grant 12-EUCLID12-0004, the Nancy Grace Roman Telescope, and JPL, which is run by Caltech under a contract for NASA. RP was supported by the Polish National Agency for Academic Exchange via Polish Returns 2019 grant. DM acknowledges support by the European Research Council (ERC) under the European Union’s FP7 Programme, GrantNo. 833031. This research has made use of the NASA Exoplanet Archive, which is operated by the California Institute of Technology, under contract with the National Aeronautics and Space Administration under the Exoplanet Exploration Program.
\end{acknowledgements}

\bibliographystyle{aasjournal}  
\bibliography{RomanAndEuclid.bib}



\end{document}